\documentclass[a4paper,11pt]{article}
\usepackage{amssymb,graphicx,amsmath}

\makeatletter 
\renewcommand\section{\@startsection{section}{1}{\z@}              {-3.25ex\@plus -1ex \@minus -.2ex}
{1.5ex \@plus .2ex}                                    {\reset@font\Large\bfseries}}
\renewcommand\subsection{\@startsection{subsection}{2}{\z@}
{3.25ex \@plus 1ex \@minus.2ex}                                    {-1em}
{\reset@font\large\bfseries}}
\@addtoreset{equation}{section} \makeatother

\input{tcilatex}
\begin{document}

\setcounter{page}{0} \topmargin0pt \oddsidemargin0mm \renewcommand{%
\thefootnote}{\fnsymbol{footnote}} \newpage \setcounter{page}{0}
\begin{titlepage}

\vspace{0.2cm}
\begin{center}
{\Large {\bf Form factors of integrable Heisenberg (higher) spin
chains}}

\vspace{0.8cm} {\large  $\text{O.~A.~Castro-Alvaredo}^{\bullet}$ and
$\text{J.~M.~Maillet}^{\circ} $}

\vspace{0.2cm}
{$^{\bullet}$ Centre for Mathematical Science, City University London, \\
Northampton Square, London EC1V 0HB, UK}\\
{$^{\circ}$ Laboratoire de Physique, ENS Lyon,  UMR 5672 du CNRS, \\
46 All\'ee d'Italie, 69364 Lyon, France}
\end{center}
\vspace{0.5cm}

\renewcommand{\thefootnote}{\arabic{footnote}}
\setcounter{footnote}{0}

\begin{abstract}
\noindent We present determinant formulae for the form factors of
spin operators of general integrable $XXX$ Heisenberg spin chains
for arbitrary (finite dimensional) spin representations. The results
apply to any ``mixed" spin chains, such as alternating spin chains,
or to spin chains with magnetic impurities.

\medskip
\medskip


\end{abstract}
\vfill{ \hspace*{-9mm}
\begin{tabular}{l}
\rule{6 cm}{0.05 mm}\\
$^\bullet \text{o.castro-alvaredo@city.ac.uk}$\\
$^\circ \text{maillet@ens-lyon.fr}$\\
\end{tabular}}
\end{titlepage}
\newpage
\section{Introduction}

Correlation functions are the main quantities characterizing quantum
theories. From an experimental point of view, measurable physical
quantities are directly related to dynamical correlation functions
(or equivalently to their Fourier transform, the structure
functions), for example in neutron scattering experiments on
ferromagnetic crystals \cite{VanHove1954,VanHove1954a}. In the
context of two dimensional integrable models various approaches to
the computation of correlation functions and form factors have been
developed over the years. It started first with models equivalent to
free Fermions for which considerable works have been necessary to
obtain full answer (see e.g.
\cite{LieM66L,Ons44,LieSM61,McC68,WuMTB76,McCTW77a,SatMJ78m}). Going
beyond the free Fermion case has been a major challenge for the last
twenty years.

For integrable massive 1+1-dimensional quantum field theories,
form factors are accessible (using some hypothesis) from the
bootstrap form factor program pioneered in the late 70's
\cite{Weisz,KW,smirnovbook}; the analytical summation of their
series corresponding to the correlation functions of local
operators remains however an open question, although accurate
numerical results exists for many theories and correlation
functions.

For integrable quantum spin chains
\cite{korepinbook,Boos:2006ux,Kitanine:2005pu} and lattice models
\cite{baxterbook}, the first attempts to go beyond free Fermion
models relied on the  Bethe ansatz techniques \cite{FST,Betherev2}
and was undertaken by  A.~G. Izergin and V.~E. Korepin (see e.g.
\cite{korepinbook} and references therein). Their approach yields
formulae for the correlation functions
\cite{korepinbook,Iz1,Iz2,Essler:1994se} written as vacuum
expectation values of some determinants depending on so-called
``dual fields" which were introduced to overcome the huge
combinatorial sums arising in particular from the action of local
operators on Bethe states. However these formulae are not completely
explicit, since these ``dual fields" cannot be eliminated from the
final result.

In the last fifteen years, two main approaches to a more explicit
computation of form factors and correlation functions have been
developed, mainly for lattice models.

One of these approaches was initiated by M. Jimbo, T. Miwa and
their collaborators \cite{JMi2,JMi,JMi3} and enables, using some
hypothesis, to compute form factors and correlation functions of
quantum spin chains of infinite length (and in their massive
regime) by expressing them in terms of traces of $q$-deformed
vertex operators over an irreducible highest weight representation
of the corresponding quantum affine algebra. These traces turn out
to satisfy an axiomatic system of equations called q-deformed
Knizhnik-Zamolodchikov (q-KZ) equations, the solutions of which
can be expressed in terms of multiple integral formulae. Using
these equations similar formulae can be obtained in the massless
regime. Recently, a more algebraic representation for the solution
of these q-deformed Knizhnik-Zamolodchikov equations have been
obtained for the XXX and XXZ (and conjectured for the XYZ) spin
1/2 chains; in these representations, all elementary blocks of the
correlation functions can be expressed in terms of some
transcendental functions
\cite{Boos:2004zq,Boos:2004tp,Boos:2005pr}. A detailed review of
the approach  can be found in \cite{Boos:2006ux}.

The second approach  has been developed by N. Kitanine, J. M.
Maillet and V. Terras \cite{JM0,JM1,KMT}. It combines the algebraic
Bethe ansatz techniques \cite{FST,Betherev2} with the solution of
the so-called quantum inverse scattering problem \cite{JM0,JM1}. It
leads in particular  to explicit determinant formulae for form
factors of the finite Heisenberg spin 1/2 XXX and XXZ chains and to
their correlation functions as explicit multiple sums. The solution
of the inverse scattering problem means in practice finding  an
explicit realization of the local operators of a large variety of
quantum spin chains in terms of the quantum monodromy matrix entries
appearing in the algebraic Bethe ansatz framework and containing in
particular  the creation operators for Bethe eigenstates of the
chain. Hence the computation of elementary blocks of the correlation
functions reduces to a soluble algebraic problem in the Yang-Baxter
algebra generated by the monodromy matrix entries. Elementary blocks
of correlation functions of the infinite XXZ spin 1/2 chain both in
the massive and massless regime as well as in the presence of a
magnetic field \cite{KMT} have been computed in terms of multiple
integrals. At zero magnetic field it gives a complete proof of the
multiple integral representations obtained in \cite{JMi2,JMi3} both
for massive and massless regimes. Hence, together with the works
\cite{JMi2,JMi3}, it also gives a proof that correlation functions
of the XXZ (inhomogeneous) chain indeed satisfy (reduced) q-deformed
Knizhnik-Zamolodchikov equations.  In addition this method has
proven to be effective in dealing with spin-spin correlation
functions in the presence of a magnetic field \cite{KMSTa,KMSTb},
dynamical correlation functions \cite{dynamical} and at non zero
temperature \cite{goehmann}, cases which were out of reach of the
vertex operator method. A recent review of this approach can be
found in \cite{Kitanine:2005pu}.

So far, we have only mentioned various works dealing with spin 1/2
chains. However, considerable effort has been made to extend the
results mentioned above to the  higher spin integrable chains
\cite{KRS,Tak82,Babu82,Babu83}. Integral formulae for the
correlation functions of the spin 1 XXZ chain have been obtained
in \cite{Idzumi:1993sp,Idzumi:1993ds,Bougourzi:1993mn}. The
correlations of spin chains of arbitrary spin have also been
studied: integral formulae for the form factors of local operators
of the XXZ chain were obtained in \cite{Konno:1994gt} and for the
correlation functions of the XYZ chain in \cite{Kojima:2005qz}.
From the Bethe ansatz method, the solution of the quantum inverse
scattering problem has been given also for higher spin cases in
\cite{JM1} leading in \cite{K} to integral formulae for the
correlation functions of the spin 1 XXX spin chain.

The aim of the present article is to employ the Bethe ansatz
approach and the solution of the quantum inverse scattering
problem for higher spins to obtain determinant formulae for the
form factors of spin operators of general integrable Heisenberg
quantum spin chains in arbitrary (mixed) finite dimensional spin
representations. More precisely, we will consider integrable
Heisenberg spin chains with different finite dimensional spin
representations at each site. With regard to their physical
properties (in particular the characteristics of the zero
temperature ground state), these ``mixed" spin chains can be
subdivided into two groups: those where spins are mixed in fixed
proportions and those where a particular type of spin dominates
while other spins can be regarded as impurities. Examples of the
first type of chains are alternating spin chains. The second type
are spin chains with magnetic impurities. The eigenstates,
spectrum and thermodynamic properties of both classes of models
have been studied within the Bethe ansatz framework. The
alternating spin 1- spin 1/2 chain has been studied in
\cite{dVW,deVega:1993sw,Pati:1997}. More general combinations of
spins have been dealt with in \cite{alt,BD}. The spin 1/2 chain
with one spin $s$ impurity has been studied in \cite{xxxs} and
especially in the context of the Kondo problem
\cite{kondo0,kondo1}. The more general situation of a spin $s$ XXX
chain with one spin $s'$ impurity was analyzed in \cite{S1} (the
special case $s=1$ has been studied in \cite{S3,rommer2}). The
effect of impurities in Heisenberg quantum spin chains has also
been the object of experimental investigation, as shown for
example in  \cite{Vajk:2002my}.

In this article we will compute the form factors of local spin
operators for the general integrable XXX (mixed) spin chains and
express our formulae in terms of determinants of elementary
functions, similar to those found in \cite{JM0} for the spin 1/2
case.

This paper is organized as follows: In section 2 we review the
algebraic Bethe ansatz framework for the general XXX  quantum spin
chains. In section 3 we derive closed formulae for all form
factors of spin operators of the XXX chain in arbitrary spin
representations. We employ our formulae to compute the total
magnetization of the chain. In section 4 we give our conclusions
and discuss some perspectives. Some lengthy computations are given
in appendices: appendix A contains proofs of two identities
involving the higher spin eigenvalues of the transfer matrix which
we have employed in our form factor computations and appendix B
presents an alternative formula for the solution of the quantum
inverse scattering problem for local spin operators useful in form
factor computations.

\section{Algebraic Bethe ansatz for XXX quantum spin chains}
\label{introduction} In this paper we will consider general
integrable XXX quantum spin chains of length $N$ and periodic
boundary conditions $S^{\alpha}_{N+i}=S^{\alpha}_i$ for
$\alpha=z,\pm$. From the algebraic Bethe ansatz view point, the
fundamental object characterizing such chains is the quantum
monodromy matrix,
\begin{eqnarray}
    T_{0;1\ldots N}^{(\frac{1}{2})}(\lambda; \{\xi\})&=& R_{0
    N}^{(\frac{1}{2},s_N)}(\lambda - \xi_{N})
     \cdots R_{0 j}^{(\frac{1}{2},s_j)}(\lambda-\xi_{j})
     \cdots R_{0 1}^{(\frac{1}{2},s_1)}(\lambda-\xi_{1})\nonumber \\
     &=&\left(%
\begin{array}{cc}
  A(\lambda; \{\xi\}) & B(\lambda; \{\xi\}) \\
  C(\lambda; \{\xi\}) & D(\lambda; \{\xi\}) \\
\end{array}%
\right)_{\!\!0}, \label{Timp}
\end{eqnarray}
given as a tensor product of $R$-matrices solutions of the
Yang-Baxter equations \cite{Yang,Baxter} which express the quantum
integrability of the system,
\begin{equation}
R_{0 j}^{(\frac{1}{2},s_j)}(\lambda-\xi_{j}) \in
\text{End}(V_{0}^{(\frac{1}{2})} \otimes V_{j}^{(s_j)}),
\end{equation}
where $V^{(s)}$ are (2s+1)-dimensional vector spaces, usually chosen
as $\mathbb{C}^{2s+1}$. The quantities $\{\xi\}=\{\xi_1, \ldots,
\xi_N\}$ are arbitrary inhomogeneity parameters attached to the
sites of the chain. The spins $s_1, \ldots, s_N$ at each site of the
chain are belonging to arbitrary finite dimensional representations
in (\ref{Timp}) which can be a priori different for different sites
of the chain. The auxiliary space, indicated by the $0$ index, has
been chosen to be 2-dimensional, which allows us to rewrite
$T(\lambda)$ in the standard matrix form with operator entries
$\{A(\lambda), B(\lambda), C(\lambda)$, $D(\lambda)\}$\footnote{We
will usually not make explicit the dependence of the operators $A$,
$B$, $C$ and $D$ and the monodromy matrix (\ref{Timp}) on the
inhomogeneity parameters $\{\xi\}$.}. As a consequence of the
Yang-Baxter equations for the $R$-matrices, these operators satisfy
a Yang-Baxter quadratic algebra and the traces of the above
constructed monodromy matrices, namely the transfer matrices,
\begin{equation}
  t^{(1/2)}(\mu)=(A+D)(\mu), \label{t12}
\end{equation}
commute among themselves for arbitrary spectral parameters $\mu,
\lambda$, namely,
\begin{equation}
  [ t^{(1/2)}(\mu) , t^{(1/2)}(\lambda)] = 0, \label{tt}
\end{equation}
leading to a large commutative subalgebra. Thus, the above
constructed monodromy matrix describes a general XXX integrable
quantum chain having different (finite dimensional) spin
representations at each site of chain.

A particular example of $R$-matrices is the one corresponding to the
spin 1/2 representation, the well-known $4 \times 4$ matrix
\begin{equation}
R^{(\frac{1}{2},\frac{1}{2})}(\lambda)=\left(
\begin{array}{cccc}
1 & 0 & 0 & 0 \\
0 &  b(\lambda) & c(\lambda) & 0 \\
0 &  c(\lambda) &  b(\lambda)& 0 \\
0 & 0 & 0 & 1  \end{array} \right), \label{r11}
\end{equation}
\noindent where the functions $b(\lambda)$ and $c(\lambda)$ are
\begin{equation}
b(\lambda)=\frac{{\lambda}}{{\lambda+ \eta}} \quad \text{and}
\quad c(\lambda)=\frac{{\eta}}{{\lambda + \eta}}. \label{bc}
\end{equation}
and $\eta$ is an arbitrary parameter. The matrix (\ref{r11}) is
normalized in such a way that the following unitarity condition is
automatically satisfied
\begin{equation}\label{uni}
R^{(\frac{1}{2},\frac{1}{2})}_{12}(\lambda)
R^{(\frac{1}{2},\frac{1}{2})}_{21}(-\lambda)=I_{12},
\end{equation}
where $I$ is the identity matrix. $R$-matrices and monodromy
matrices of higher spin representations can be constructed from
(\ref{bc}) by means of the fusion procedure developed in
\cite{KRS}. It gives,
\begin{equation}
R^{(\frac{1}{2},s)}(\lambda) = \frac{1}{ {\lambda + \eta \left(s +
\frac{1}{2}\right)}} \left(
\begin{array}{cc}
         \lambda+ \eta \left(S^{z}+\frac{1}{2}\right)  & \eta S^{-} \\
        \eta S^{+} & \lambda+ \eta \left(\frac{1}{2}- S^{z}\right)  \\
       \end{array}\right),\label{r12}
\end{equation}
where $S^{\pm}, S^{z}$ are generators of the $su(2)$ algebra in
the spin $s$ representation with standard commutation relations.

\subsection{The Bethe ansatz equations}\label{bethe}\indent \\

 \noindent Having introduced $R$- and $T$-matrices we can now
proceed to characterize the energy eigenstates of the chain. The key
observation there is that the local Hamiltonian operator can be
identified as a simple function of the transfer matrix. Hence
finding its spectrum reduces to constructing the common spectrum of
transfer matrices $t^{(1/2)}(\mu)$ for arbitrary values of the
spectral parameter $\mu$ (since they all commute). The algebraic
Bethe ansatz method starts by constructing a particular eigenstate
of the transfer matrix, called the
 reference state $|0\rangle$, which is annihilated by the operator
$C(\lambda)$. In order to identify such a state we notice that for
every vector space $V_j$ one can find a vector $|0\rangle_{j}$ of
dimension $2s_j+1$ which fulfils the following equality
\begin{equation}
R_{0 j}^{(\frac{1}{2}, s_j)}(\lambda-\xi_j) |0\rangle_{j} = \left(%
\begin{array}{cc}
  1 & * \\
  0 & \frac{\lambda - \xi_j^{-} -s_j \eta}{\lambda  - \xi_j^{-} +s_j \eta} \\
\end{array}%
\right)|0\rangle_{j}.
\end{equation}
\noindent Here, and in the following, we use the notation
$u^{\pm}= u \pm \eta/2$ for any complex number $u$. The reference
state $|0\rangle$ is nothing but the tensor product
\begin{equation}\label{vac}
 |0\rangle= \bigotimes_{j=1}^{N} \, |0\rangle_{j} \qquad \text{with}
\qquad  |0\rangle_{j}= \left(
\begin{array}{c}
  1 \\
  0 \\
  \vdots\\
  0
\end{array}
\right)_{j},
\end{equation}
\noindent that is a completely ferromagnetic state with all spins
up. The monodromy matrix (\ref{Timp}) acts on the reference state
as
\begin{equation}
T_{0;1\ldots N}^{(1/2)}(\lambda)  |0\rangle=\left(%
\begin{array}{cc}
  1 & * \\
  0 & d(\lambda) \\
\end{array}%
\right) |0\rangle,
\end{equation}
\noindent namely
\begin{equation}
    A(\lambda)  |0\rangle = |0\rangle, \quad C(\lambda)|0\rangle=0 \quad \text{and}
    \quad D(\lambda)  |0\rangle= d(\lambda) |0\rangle,
\end{equation}
\noindent with \begin{eqnarray}\label{d}
    d(\lambda)&=&
    \prod_{j=1}^{N} \frac{\lambda-\xi_{j}^{-} - s_j \eta}{\lambda-\xi_{j}^{-} + s_j
    \eta}=
    \prod_{j=1}^{N}\left[\prod_{k=1}^{2s_j} b(\lambda-\xi_{j}^{-}-(k-s_j)\eta)\right].
\end{eqnarray}
\noindent It follows from the definition above that $d(\lambda)$ has
zeros at $\lambda = \xi_{j}^{-} + s_j \eta$ , and poles at $\lambda
= \xi_{j}^{-} - s_j \eta$ for all $j=1,\ldots, N$. Let us now
proceed to the description of the spectrum. Eigenvectors of the
transfer matrix other than $|0\rangle$ will be constructed within
the Bethe ansatz framework by acting successively on $|0\rangle$
with operators $B(\lambda_i)$. In general such eigenvectors will be
of the form
\begin{equation}
\left|\Psi(\{\lambda\})\right\rangle=B(\lambda_{1})\ldots
B(\lambda_{\ell}) |0\rangle. \label{bs}
\end{equation}
\noindent We want to select out those states (\ref{bs}) which are
eigenstates of the transfer matrix and hence of the Hamiltonian. The
condition that (\ref{bs}) is a common eigenstate of (\ref{t12}) for
all values of $\mu$ is a set of coupled algebraic equations for the
spectral parameters $\lambda_{i}$ describing the state (\ref{bs})
which are known as Bethe ansatz equations
\begin{equation}
\prod_{k=1}^{\ell}
\frac{b(\lambda_{j}-\lambda_{k})}{b(\lambda_{k}-\lambda_{j})}=
-d(\lambda_{j}), \label{ba}
\end{equation}
\noindent The eigenvalues of $t^{(1/2)}(\mu)$ corresponding to the
eigenvectors (\ref{bs}) fulfilling (\ref{ba}) are given by
\begin{equation}\label{eig}
    \tau^{(1/2)}(\mu,\{\lambda\})= \prod_{j=1}^{\ell}
    b^{-1}(\lambda_{j}- \mu) + d(\mu) \prod_{j=1}^{\ell}
    b^{-1}(\mu-\lambda_{j}).
\end{equation}

\subsection{Fusion} \indent \\\label{fusion}

\noindent The mechanism of fusion for quantum spin chains was
first developed in \cite{KRS} for the XXX spin chain (and later on
in \cite{KR} for the XXZ spin chain). It provides a procedure for
constructing higher spin objects ($R$-, monodromy and transfer
matrices) starting with spin 1/2 objects. The fusion identities
for the quantum monodromy matrix can be written as follows
\begin{eqnarray}
P_{12}T_{1}^{(\frac{{1}}{2})}(x^{-}+s\eta)
T_{2}^{(s-\frac{{1}}{2})}(x^{-})P_{12}
=\left(\!\!\begin{array}{cc}
T_{\langle 12\rangle}^{(s)}(x) & 0 \\
* & \chi(x+(s-1)\eta) T_{( 12 )}^{(s-1)}(x-\eta) \\
\end{array}\!\!
\right), \label{dttd}
\end{eqnarray}
where for the sake of simplicity we  dropped the quantum indices
$1,\ldots, N$ of the monodromy matrices (\ref{Timp}) and the
monodromy matrix associated to the spin $s$ in the auxiliary
(first) space can be computed as a product over $R^{(s,s_j)}$
matrices as,
\begin{eqnarray}
    T_{0;1\ldots N}^{(s)}(\lambda; \{\xi\})&=& R_{0
    N}^{(s,s_N)}(\lambda - \xi_{N})
     \cdots R_{0 j}^{(s,s_j)}(\lambda-\xi_{j})
     \cdots R_{0 1}^{(s,s_1)}(\lambda-\xi_{1})\nonumber \\
     \label{Timps}
\end{eqnarray}
with,
\begin{equation}
R_{0 j}^{(s,s_j)}(\lambda-\xi_{j}) \in \text{End}(V_{0}^{(s)}
\otimes V_{j}^{(s_j)}).
\end{equation}
The matrix
\begin{equation}
P_{12}= P^{+}_{\langle 12\rangle} \oplus P^{-}_{(12)}\,,
\end{equation}
is a direct sum of projectors $P^{\pm}$ onto the vector spaces $
V_{\langle 12\rangle}^{(s)}\sim \mathbb{C}^{2s+1}$ and
$V_{(12)}^{(s-1)}\sim \mathbb{C}^{2s-1}$ resulting from the tensor
decomposition
\begin{equation}
    V_{1}^{(s-\frac{1}{2})} \otimes V_{2}^{(\frac{1}{2})} \simeq V_{\langle 12\rangle}^{(s)}\oplus V_{(12)}^{(s-1)},
\end{equation}
with $V_{0}^{(\frac{1}{2})}\sim \mathbb{C}^{2}$ and
$V_{\hat{0}}^{(s-\frac{1}{2})}\sim \mathbb{C}^{2s}$. Further
\begin{equation}
   \chi(u)=A(u^{+}) D(u^{-})- B(u^{+})C(u^{-}),\label{qd}
\end{equation}
is the quantum determinant of the transfer matrix (\ref{Timp}),
which commutes with all operators $A,B,C,D$. A completely analog
expression holds when replacing $T$-matrices by $R$-matrices which
permits the construction of (\ref{r12}) starting with (\ref{r11}).
A direct consequence of (\ref{dttd}) is the following recursive
relation involving transfer matrices of various spins,
\begin{eqnarray}
    t^{(s)}(x)= t^{(1/2)}(x^{-}+s\eta)
    t^{(s-1/2)}(x^{-})-\chi(x+(s-1)\eta) t^{(s-1)}(x-\eta),
    \label{ide1}
\end{eqnarray}
and therefore,
\begin{eqnarray}
    \tau^{(s)}(x)=\tau^{(1/2)}(x^{-}+s\eta)
    \tau^{(s-1/2)}(x^{-})-d(x^{-}+(s-1)\eta) \tau^{(s-1)}(x-\eta),
    \label{eigen}
\end{eqnarray}
for the corresponding eigenvalues on a Bethe state
$|\Psi(\{\lambda\})\rangle$ and where for simplicity we have
dropped the explicit dependency of the eigenvalues $\tau$ on the
spectral parameters $\lambda$ of the Bethe state. In \cite{KR} it
was also realized that, as a consequence of (\ref{ide1}), a
generating functional for monodromy matrices of higher spin could
be constructed as
\begin{equation}
   F(z)= \frac{1}{1-z t^{(1/2)}(u)+z^2 \chi(u^{+})}=
    \sum_{k=0}^{\infty} z^{k} t^{(k/2)}(u^{-}+ k\eta/2).
    \label{genera}
\end{equation}
Here we take  $t^{(0)}(x)=1$ for any values of $x$,  and define
$z$ as a shift operator which acts on an operator $t(x)$ as
\begin{equation}
   t(x)z=z t(x+\eta).
\end{equation}
There are some differences between (\ref{genera}) and the formula
given in \cite{KR}, which are due to the different normalization
chosen for the $R$-matrices in this paper. With the  help of
(\ref{ide1}) we can easily see that indeed (\ref{genera})
generates any higher spin monodromy matrices. One must only match
the terms on the l.h.s. and r.h.s. which contain the same powers
of  $z$, paying special attention to the fact that $z$ is a shift
operator. For example, expanding the l.h.s. of (\ref{genera}) we
find the following first terms
\begin{eqnarray}
&&F(z)= 1 + z t^{(1/2)}(u) + \left[z t^{(1/2)}(u)z
t^{(1/2)}(u)-z^{2} \chi(u^{+})\right] + \ldots
\end{eqnarray}
The $\mathcal{O}(1)$ and $\mathcal{O}(z)$ terms match trivially
those on the r.h.s. of (\ref{genera}) whereas the
$\mathcal{O}(z^{2})$ term is
\begin{equation}
\left[z t^{(1/2)}(u)\right]^{2}-z^{2} \chi(u^{+})=
z^{2}\left[t^{(1/2)}(u+\eta) t^{(1/2)}(u) -
\chi(u^{+})\right]=z^{2} t^{(1)}(u^{+}),
\end{equation}
which exactly agrees with (\ref{ide1}) when taking $s=1$.
Similarly, one can establish the agreement with (\ref{ide1}) at
every order in $z$.

The existence of the generating function $F(z)$ translates into
the existence of a generating function for the eigenvalues of the
operators $t^{(s)}(u)$. They appear as coefficients of the
corresponding Laurent series in $z$ for the generating function.
This allowed the authors of \cite{KR} to find explicit formulae
for the eigenvalues which, translated into our present
normalization, take the form
\begin{equation}
    \tau^{(s)}(u, \{\lambda\})=\sum_{\alpha=0}^{2s}
    C_{\alpha}^{(s)}(u) \prod_{p=1}^{\ell} \frac{(u^{+}-\lambda_p+
    s\eta)
    (u^{-}-\lambda_p -s\eta)}{(u^{+}-\lambda_p+ (\alpha-s) \eta)
    (u^{-}-\lambda_p +(\alpha-s)\eta)}, \label{ex}
\end{equation}
with
\begin{equation}\label{C}
 C_{\alpha}^{(s)}(u) = \prod_{k=\alpha}^{2s-1} d(u^{+}+ (k-s) \eta) \qquad
 \text{and}\qquad C_{2s}^{(s)}(u)=1.
\end{equation}
It is now easy to check that (\ref{ex}) indeed solves
(\ref{eigen}). Similar relations hold for the XXZ case as well.

\subsection{The inverse scattering problem}\indent\\

\noindent In the context of the algebraic Bethe ansatz approach,
the inverse scattering problem is understood as the problem of
expressing quantum local operators of the chain in terms of the
operators $A, B, C, D$ with generate the Bethe states. The
solution to this problem for a large variety of spin chains was
provided in \cite{JM0,JM1}. In particular, the reconstruction
formulae for the spin operators of the XXX chain in an arbitrary
spin $s_j$ representation were found to be
\begin{eqnarray}
  S_j^{\alpha} &=& \left[\prod_{k=1}^{j-1} t^{(s_k)}(\xi_k)\right]
  \Lambda_{\alpha}^{(s_j)}(\xi_j) \left[\prod_{k=1}^{j} t^{(s_k)}
  (\xi_k)^{-1}\right],\qquad \alpha =\pm ,z\label{sp}
\end{eqnarray}
with
\begin{eqnarray}
\Lambda _{\alpha }^{(s)}(u)&:=&\text{Tr}_{0}\left( S_{0}^{\alpha
}T_{0}^{(s)}(u)\right) \nonumber
\\ &=&\sum_{k=1}^{2s}t^{(s-\frac{k}{2})}\left( u+\frac{k\eta
}{2}\right) \Lambda
_{\alpha }^{(\frac{1}{2})}(u^{-}+(k-s)\eta )t^{(\frac{k-1}{2})}\left( u^{-}+\frac{%
(k-2s)\eta }{2}\right)\label{L1} \\
&=&\sum_{k=1}^{2s}t^{(\frac{k-1}{2})}\left( u^{-}+\frac{%
(k-2s)\eta }{2}\right)\Lambda _{\alpha
}^{(\frac{1}{2})}(u^{-}+(k-s)\eta )t^{(s-\frac{k}{2})}\left(
u+\frac{k\eta }{2}\right)\!\!.  \label{Lam2}
\end{eqnarray}
The expression (\ref{L1}) was derived in \cite{JM1} using
(\ref{dttd}), whereas the non-trivial equivalence between the
expressions (\ref{L1}) and (\ref{Lam2}) is established in appendix B
of the present manuscript. The existence of these two expressions
will be very useful for later form factor computations.

\section{Form factors}
Let us start by introducing several formulae which we will use in
the course of our computations. First of all, we will need the
action of the operators $A$ and $D$ on a generic quantum state
$\left|\Psi(\{\lambda\})\right\rangle:=\prod_{k=1}^{\ell}
B(\lambda_k) \left|0 \right \rangle$,
\begin{eqnarray}
  A(x) \left|\Psi(\{\lambda\})\right\rangle &=& \prod_{k=1}^{\ell}
   b^{-1}(\lambda_k-x) \left|\Psi(\{\lambda\})\right\rangle\nonumber \\
   &-&\sum_{p=1}^\ell c({\lambda_p - x-\eta})\prod_{k\neq
p} b^{-1}(\lambda_k-\lambda_p) B(x) \prod_{k\neq
p}B(\lambda_k) \left|0\right\rangle, \label{a}\\
D(x) \left|\Psi(\{\lambda\})\right\rangle
&=&d(x)\prod_{k=1}^{\ell}
   b^{-1}(x-\lambda_k) \left|\Psi(\{\lambda\})\right\rangle \nonumber \\
   &+&\sum_{p=1}^\ell d(\lambda_p)c(\lambda_p - x-\eta)\prod_{k\neq
p} b^{-1}(\lambda_p-\lambda_k) B(x) \prod_{k\neq p}B(\lambda_k)
\left|0\right\rangle.\label{dd}
\end{eqnarray}
We see that this action is divided into two kinds of terms, which
are traditionally  refer to as ``direct" and ``indirect" terms.
Direct terms are those which leave the original state unchanged up
to a scalar factor while in indirect terms one of the parameters
$\lambda_p$ has been replaced by the parameter $x$.

Secondly, we need to compute scalar products of quantum states,
either  of two Bethe states or of one Bethe and one generic state.
The formula for the scalar product of two Bethe states was
originally obtained in \cite{Kor,Iz1}. Later on it was proven
\cite{ns} that a completely analog formula also holds for the
scalar product of a Bethe state and an arbitrary state. Finally,
the same formula was re-derived in \cite{JM0} with the help of the
$F$-basis introduced in \cite{SSM}. It takes the following form,
\begin{eqnarray}
   \langle \psi(\{\mu\})| \psi(\{\lambda\})
   \rangle :=S_{\ell}(\{\mu\}, \{\lambda\})
   = \frac{\det H(\{\mu\},\{\lambda\})}{\prod\limits_{i<j}(\lambda_i - \lambda_j)
   (\mu_j-\mu_i)}= S_{\ell}(\{\lambda\}, \{\mu\}) ,\label{spr}
\end{eqnarray}
where $H(\{\mu\},\{\lambda\})$ is a $\ell \times \ell$ matrix of
components
\begin{equation}\label{Hab}
    H_{ab}=\frac{\eta}{\mu_a - \lambda_b} \left[ \prod_{i \neq a}
     (\mu_i - \lambda_b + \eta) -
    d(\lambda_b) \prod_{i \neq a}(\mu_i - \lambda_b - \eta)
    \right],
\end{equation}
and $\{\mu\}=\{\mu_1, \ldots, \mu_{\ell}\}$ and
$\{\lambda\}=\{\lambda_1, \ldots, \lambda_{\ell}\}$
 are a Bethe state and an
arbitrary state, respectively.

Finally, we would like to recall a property of determinants which
we will employ in the computation of the magnetization below:
given two $\ell \times \ell$ matrices $\mathcal{X}$ and
$\mathcal{Y}$ such that all rows of $\mathcal{Y}$ are identical
(rank 1 matrix) the following equality holds,
\begin{equation}
    \det(\mathcal{X}+\mathcal{Y})=\det \mathcal{X} +
    \sum_{p=1}^{\ell} \det \mathcal{X}^{(p)}, \label{xy}
\end{equation}
where
\begin{eqnarray}
\mathcal{X}^{(p)}_{ab}&=&\ \mathcal{X}_{ab}\qquad
\text{for}\quad b\neq p,\\
\mathcal{X}^{(p)}_{ap}&=& \mathcal{Y}_{ap}.
\end{eqnarray}

\subsection{The form factors of $S_j^{z}$} \label{sz}\indent\\\\
 We define the non-vanishing form factors of the
local operator $S^{z}_{j}$ in the spin $s_j$ representation as
\begin{eqnarray}
  F^{z}_{\ell}(j,\{\mu\}, \{\lambda\}) &=& \left\langle \psi (\{\mu\})\right|
  S^{z}_{j}\left|\psi
  (\{\lambda\})\right\rangle,
  \label{form}
\end{eqnarray}
with $\{\mu\}$ and $\{\lambda\}$ being two sets of $\ell$ Bethe
roots, therefore characterizing two Bethe states.

Inserting (\ref{sp}) with (\ref{L1}) and $\alpha=z$ into
(\ref{form}) we obtain the following sum of matrix elements
\begin{eqnarray}
  &&F_{\ell}^{z}(j,\{\mu\}, \{\lambda\}) =
  \frac{\phi_{j-1}(\{\mu\})}{2 \phi_{j}(\{\lambda\})}
   \sum_{k=1}^{2s_j} \left[\tau^{(s_j-\frac{k}{2})}\left(\xi_j + \frac{{k\eta}}{2},\{\mu\}\right)
  \right. \nonumber \\
   &&\left. \times
   \left\langle \psi (\{\mu\})\right|
   (A-D)(\nu_j(k))t^{(\frac{k-1}{2})}\left(\xi_j^{-} +
   {\frac{(k-2s_j)\eta}{2}}\right)
   \left|\psi(\{\lambda\})\right\rangle\right],
  \label{form2}
\end{eqnarray}
where the functions
\begin{equation}
\phi_{j}(\{\mu\})=\prod_{k=1}^{j} \tau^{(s_k)}(\xi_k,\{\mu\}),
\end{equation}
originate from the actions
\begin{eqnarray}
  \langle \psi(\{\mu\})|
   \prod_{k=1}^{j-1}t^{(r_k)}(\xi_k)&=&
   \phi_{j-1}(\{\mu\})\langle \psi(\{\mu\})|,\\
\prod_{k=1}^{j}t^{(r_k)}(\xi_k)^{-1}
 \left|\Psi(\{\lambda\})\right\rangle &=& \phi_{j}(\{\lambda\})^{-1}
\left|\Psi(\{\lambda\})\right\rangle, \label{phi}
\end{eqnarray}
and we introduced the variable
\begin{equation}\label{nu}
    \nu_j(k)=\xi^{-}_j+(k-s_j)\eta.
\end{equation}
In (\ref{form2}) we have not yet replaced the operators
$t^{((k-1)/2)}$ by their eigenvalues on the Bethe state
$|\Psi\{\lambda\}\rangle$ for reasons which will become apparent
below. Rewriting now $A-D=2A-(A+D)$ and recalling the general
action of the operator $A$ on a Bethe state given in (\ref{a}) we
find that (\ref{form2}) is equivalent to
\begin{eqnarray}
  && F_{\ell}^{z}(j,\{\mu\}, \{\lambda\}) =
  \frac{\phi_{j-1}(\{\mu\})}{2 \phi_{j}(\{\lambda\})} \left[g(j,\{\mu\})
  S_{\ell}(\{\mu\},\{\lambda\})  \right.\label{form22}\\
  && \left. - 2\sum_{k=1}^{2s_j} \sum_{p=1}^{\ell}
  \left[\tau^{(s_j-\frac{k}{2})}\left(\xi_j +
  \frac{{k\eta}}{2},\{\mu\}\right)
  \tau^{(\frac{k-1}{2})}\left(\xi_j^{-} + \frac{{(k-2s_j)\eta}}{2},\{\lambda\}\right)
\right.\right.\nonumber\\
 && \times \left. \left. \frac{\eta}{\mu_p-\nu_j(k)}
 \prod_{k\neq p} b^{-1}(\mu_k-\mu_p)
S_{\ell}(\{\lambda\}, \{\mu, {\mu}_p \rightarrow
\nu_j(k)\})\right]\right].\nonumber
\end{eqnarray}
Here  $g(j,\{\mu\})$ is the function defined in equation
(\ref{for1}) of appendix A. The term proportional to
$g(j,\{\mu\})$ collects all contributions which are ``direct" in
the sense indicated after (\ref{a})-(\ref{dd}). The form of this
term results from acting with both operators $t^{(s_j-k/2)}$ and
$t^{(k-1)/2}$ on the Bethe state $|\Psi(\{\mu\})\rangle$. By
carrying out the computation in this way we achieve that the
direct contribution to the form factor is proportional to the
function $g(j,\{\mu\})$ and we can take advantage of the identity
$(\ref{fa})$ proven in appendix A.

The matrix element $S(\{\lambda\},\{\mu, {\mu}_p \rightarrow
\nu_j(k)\})$ in (\ref{form22}) is the scalar product (\ref{spr})
with the parameter $\mu_p$ replaced by $\nu_j(k)$. It can be
written as
\begin{equation}
S(\{\mu, {\mu}_p \rightarrow x\} ,\{\lambda\})=\prod_{j\neq p
}\frac{(\mu_p-\mu_j)}{(x-\mu_j)}
 \frac{\det H^{(p)}(x)}{\prod\limits_{i<j}(\lambda_i - \lambda_j)
   (\mu_j-\mu_i)},\label{pa2}
\end{equation}
where $H^{(p)}(x)$ is a matrix such that
$H^{(p)}_{ab}=H_{ab}(\{\lambda\},\{\mu\})$ for $b\neq p$ and
\begin{eqnarray}
   H^{(p)}_{ap}(x)&=& \frac{\eta}{\lambda_a - x}
   \left[ \prod_{i \neq a} (\lambda_i - x + \eta) -
    d(x) \prod_{i \neq a}(\lambda_i -x - \eta)
    \right] \label{hp}\\
    &=& \frac{t(\lambda_a ,x)}{b(\lambda_a -x)}\prod_{i = 1}^{\ell} (\lambda_i - x )
    \left[\prod_{i \neq a} b^{-1}(\lambda_i-x)-
    d(x) \prod_{i \neq a}{b^{-1}(x - \lambda_i)}
    \right],\nonumber
\end{eqnarray}
with
\begin{equation} \label{tlx}
    t(\lambda,x)=\frac{\eta}{(\lambda-x)(\lambda-x +
    \eta)}=-i \partial_{\lambda} p^{(\frac{1}{2})}(\lambda-x^{-}),
\end{equation}
and $p^{(1/2)}(\lambda)$ being the momentum defined in (\ref{pp}).
Inserting (\ref{pa2}) into (\ref{form22}) with $x=\nu_j(k)$ we
obtain the following expression for the form factors
\begin{eqnarray}
F_{\ell}^{z}(j,\{\mu\},\{\lambda\}) =
\frac{\phi_{j-1}(\{\mu\})}{2\phi_{j}(\{\lambda\})}\frac{\left[{g(j,\{\mu\})
\det H-2
\sum\limits_{p=1}^{\ell}\prod\limits_{k=1}^{\ell}(\mu_k-\mu_p +
\eta)  \det \tilde{{H}}^{(p)}(\xi_j)}\right]}
{\prod\limits_{i<j}(\lambda_i - \lambda_j)
(\mu_j-\mu_i)},\label{fin2forz}
\end{eqnarray}
where
\begin{eqnarray}
 \tilde{H}^{(p)}_{ab} &=& H_{ab}(\{\lambda\},\{\mu\}) \qquad\text{for}\qquad b \neq p,\\
   \tilde{H}^{(p)}_{ap}&=&   f^{(p)}(j, \{\mu\}, \{\lambda\})
\end{eqnarray}
and $f^{(p)}(j,\{\mu\},\{\lambda\})$ and $g(j,\{\mu\})$ are the
functions introduced at the beginning of the appendix A. Thanks to
the identities (\ref{id1}) and (\ref{id2}) this expression can be
simplified to
\begin{eqnarray}
  F_{\ell}^{z}(j,\{\mu\},\{\lambda\}) = \frac{\phi_{j}(\{\mu\})}{\phi_{j}(\{\lambda\})}
 \frac{s_j \det H-
\sum\limits_{p=1}^{\ell}\prod\limits_{k=1}^{\ell}(\mu_k-\mu_p +
\eta)\det \mathcal{Z}^{(p)}(\xi_j)
  }{\prod\limits_{i<j}(\lambda_i - \lambda_j)
(\mu_j-\mu_i)},\label{finforz}
\end{eqnarray}
with
\begin{eqnarray}
  \mathcal{Z}^{(p)}_{ab} &=& H_{ab}(\{\lambda\},\{\mu\}) \qquad\text{for}\qquad b \neq p,\\
   \mathcal{Z}^{(p)}_{ap}&=& -i\partial_a p^{(s_j)}(\lambda_a-\xi_j^{-})
 \, \prod_{k=1}^{\ell}
  \frac{\lambda_k-\xi_{j}^{-}+s_j\eta}
  {\mu_k-\xi_{j}^{-}+s_j\eta}
  .
\end{eqnarray}
Notice that, since the entries $\mathcal{Z}^{(p)}_{ap}$ depend on
both indices $a, p$ we can not write (\ref{finforz}) in terms of a
single determinant by employing (\ref{xy}). However, the factor $
\prod_{k=1}^{\ell}(\mu_k-\mu_p + \eta)$ will be canceled when
normalizing by the norm of the Bethe states, which allows a
particularly simple form for the magnetization. We compute this
quantity in the next section. Let us also remark that the
computation of (\ref{form}) can be done in two different ways, which
of course must lead to the same result: in (\ref{form22}) we decided
to make the operators $(A-D)(\nu_j(k))$ act on the Bethe state
$\{\mu\}$, however we could as well have chosen to act on the Bethe
state $\{\lambda\}$. If we do that and employ the same
reconstruction formula (\ref{L1}) as before, we will obtain an
expression which looks rather different from (\ref{finforz}) and
cannot be simplified in any obvious way. However, we can
alternatively employ the reconstruction formula (\ref{Lam2}) to do
the same computation. In that case we will obtain a formula
completely analogous to (\ref{finforz}) where the roles of
$\{\lambda\}$ and $\{\mu\}$ are just exchanged.

\subsection{Magnetization}\indent\\

\noindent One interesting physical quantity we can now compute is
the magnetization at site $j$ which is defined as
\begin{equation}
  \langle S_j^{z}\rangle = \frac{F_{\ell}^{z}(j,\{\lambda\},\{\lambda\})}
  {\langle \psi(\{\lambda\})|\psi(\{\lambda\}) \rangle}, \label{s}
\end{equation}
The norm of the Bethe state  is given in terms of the Gaudin
matrix $\Phi^{\prime}(\{\lambda\})$ as follows \cite{Kor}
\begin{eqnarray}
  {\langle \psi(\{\lambda\})|\psi(\{\lambda\}) \rangle} &=&
  {\eta}^{\ell} \prod_{a\neq b}{b^{-1}(\lambda_{a}-\lambda_b)}
    \det\Phi^{\prime}(\{\lambda\}),
\end{eqnarray}
with
\begin{eqnarray}
  \Phi^{\prime}_{ab} &=& -\frac{\partial}{\partial \lambda_{b}}
  \ln \left( \frac{1}{d(\lambda_{a})}
  \prod_{\substack{k=1 \\ k \neq a}}^{\ell}
  \frac{b(\lambda_{a}-\lambda_k)}{b(\lambda_{k}-\lambda_a)}\right),
  \label{gaudin}
\end{eqnarray}
and can be obtained from the scalar product formula (\ref{spr}) in
the limit $\lambda_a \rightarrow \mu_a$ for all $a=1 \ldots \ell$.
In particular, it is easy to prove that for two identical Bethe
states, the entries of the matrix $H(\{\lambda\})$ defined in
(\ref{Hab}) become proportional to those of the Gaudin matrix as
\begin{equation}
H_{ab} = \prod_{i=1}^{\ell}(\lambda_{i}-\lambda_b+\eta)
\Phi^{\prime}_{ab}.
\end{equation}
and therefore
\begin{equation}
\label{hlimit} \det H(\{\lambda\}) =
\prod_{i,j=1}^{\ell}(\lambda_{i}-\lambda_j+\eta)\det
\Phi^{\prime}(\{\lambda\}).
\end{equation}
Using these identities and (\ref{fin2forz}) we obtain
\begin{equation}
  \langle {S}^{z}_j \rangle  =
  \frac{s_j\det \Phi^{\prime}-\sum_{a=1}^{\ell} \det \mathcal{M}^{(a)}}{\det
  \Phi^{\prime}},\label{mag}
\end{equation}
where the matrix $\mathcal{M}^{(a)}$ is such that
\begin{eqnarray}
  \mathcal{M}^{(a)}_{ij}&=& \Phi^{\prime}_{ij} \qquad \text{for} \qquad j \neq a, \\
  \mathcal{M}^{(a)}_{ia} &=& -i\partial_{\lambda_a}p^{(s_j)}(\lambda_a
  -\xi_j^{-}),\label{mm}
\end{eqnarray}
where $p^{(s_j)}(\lambda)$ is the momentum (\ref{pp}) defined in
appendix A. Since $\mathcal{M}^{(a)}_{ia}$ is independent of the
value of the index $i$, the identity (\ref{xy}) can be used to
bring (\ref{mag}) into the form
\begin{equation}
\langle \mathcal{S}^{z}_j \rangle =  s_j \frac{\det(
\Phi^{\prime}-\mathcal{M}/s_j)}{\det\Phi^{\prime}},
\end{equation}
where $\mathcal{M}$ is a rank 1 matrix whose entries are
$\mathcal{M}_{ab}= \mathcal{M}^{(b)}_{ab}$. The total
magnetization of the chain can be computed as
\begin{equation}
\mu_{\text{tot}}=\sum_{j=1}^{N}  \langle S_{j}^{z}\rangle=s_0
\frac{\det (\Phi^{\prime}- \mathcal{M}_{\text{tot}}/s_0)}{\det
\Phi^{\prime}},
\end{equation}
where $\mathcal{M}_{\text{tot}}$ is a rank 1 matrix with entries
\begin{equation}
(\mathcal{M}_{\text{tot}})_{ia}= \sum_{j=1}^{N}
-i\partial_{\lambda_a}p^{(s_j)}(\lambda_a -\xi_j^{-}),
\end{equation}
and $s_0$ is the total spin of the completely ferromagnetic state
$|0\rangle$
\begin{equation}\label{s0}
    s_0=\sum_{j=1}^{N} s_j.
\end{equation}
The total magnetization can be computed by noticing the following
property of the Gaudin matrix (\ref{gaudin})
\begin{equation}
    \sum_{b=1}^{\ell}\Phi^{\prime}_{ab}=(\mathcal{M}_{\text{tot}})_{ka}.
\end{equation}
Therefore
\begin{equation}
    \mu_{\text{tot}}=s_0 \det\left({I}- \frac{\mathcal{U}}{s_0}\right)=
    s_0-\ell,\label{ts}
\end{equation}
where ${I}$ denotes the identity matrix and $\mathcal{U}$ is the
matrix,
\begin{equation}\label{u}
    \mathcal{U}_{ab}=1 \qquad \forall \qquad a, b.
\end{equation}
This computation provides a consistency check of the general
formula (\ref{fin2forz}) as it reproduces the values of the
magnetization known for particular cases. For example, for the
pure spin 1/2 XXX chain in the ground state we have $\ell=N/2$ and
$s_0 = N/2$ so that the total magnetization vanishes (see e.g.
\cite{JM0}). Another well known example is the spin 1/2 XXX chain
with one spin $s$ impurity. In this case the ground state is also
characterized by $\ell=N/2$ roots, but the total spin of the chain
is $s_0=(N-1)/2 +s$. Therefore the total magnetization becomes
$s-1/2$, in agreement with the value computed in \cite{kondo1}.
\subsection{The form factors of $S_j^{+}$}\indent\\

\noindent In this section we compute the form factors of the
operator $S_j^{+}$ of the quantum XXX spin chain. That is a spin
generator sitting at site $j$ of the chain and living in the spin
$s_j$ representation. The only non-vanishing form factors are
\begin{eqnarray}
  F^{+}_{\ell}(j,\{\lambda\}, \{\mu\}) &=& \left\langle \psi (\{\lambda\})\right|
  S^{+}_{j}\left|\psi
  (\{\mu\})\right\rangle,
  \label{formpm}
\end{eqnarray}
where $|\Psi(\{\mu\})\rangle$ and $|\Psi(\{\lambda\})\rangle$ are
two Bethe states with $\ell+1$ and $\ell$ Bethe roots,
respectively. From (\ref{sp}) and (\ref{L1}) we obtain
  \begin{eqnarray}
  && F_{\ell}^{+}(j,\{\lambda\}, \{\mu\}) = \frac{\phi_{j-1}(\{\lambda\})}{\phi_{j}(\{\mu\})}
   \sum_{k=1}^{2s_j}\left[
   \tau^{(s_j-\frac{k}{2})}\left(\xi_j + {k\eta}/{2}, \{\lambda\}\right)\right.\nonumber \\
   && \left. \times \tau^{(\frac{k-1}{2})}\left(\xi_j^{-} + {(k-2s_j)\eta}/{2}, \{\mu\}  \right)
   S_{\ell+1}(\{\mu\},\{\lambda,\nu_j(k)\})\right],
  \label{formmin}
\end{eqnarray}
where $\nu_j(k)$ is again the variable introduced in (\ref{nu})
and $S_{\ell+1}(\{\mu\},\{\lambda,\nu_j(k) \})$ is the scalar
product introduced in (\ref{spr}) with $\{\lambda,\nu_j(k)\}$ the
set of $\ell+1$ variables $\{\lambda_1, \ldots, \lambda_\ell,
\nu_j(k)\}$ in the given order, that is $\nu_j(k)$ is the variable
number $\ell+1$ (notice that the ordering of variables is
important in (\ref{spr})). Employing the definition (\ref{spr}) we
have
\begin{eqnarray}
 S_{\ell+1}(\{\mu\},\{\lambda,\nu_j(k)\})=
 \frac{\det H(\{\mu\},\{\lambda,\nu_j(k)\})}
 {\prod\limits_{1\leq i<j\leq \ell+1} (\mu_j-\mu_i)\prod\limits_{1\leq i<j\leq \ell}
(\lambda_i-\lambda_j)}
 \prod\limits_{i=1}^{\ell}(\lambda_i-\nu_j(k))^{-1},
\end{eqnarray}
where $H$ is a matrix with components $H_{ab}$ given by
(\ref{Hab}), except for the column $\ell+1$ whose entries are
given by (\ref{hp}) with $x=\nu_j(k)$. This means that
(\ref{formmin}) can be rewritten as
\begin{eqnarray}
  && F_{\ell}^{+}(j,\{\lambda\}, \{\mu\}) = \frac{\phi_{j-1}(\{\lambda\})}{\phi_{j}(\{\mu\})}
   \frac{\det H^{+}(\xi_j)}
    {\prod\limits_{1\leq i<j\leq \ell+1}(\mu_j-\mu_i)\prod\limits_{1\leq i<j\leq \ell}
   (\lambda_i-\lambda_j)},
  \label{formminfin}
\end{eqnarray}
where
\begin{eqnarray}
  H^{+}_{ab} &=& H_{ab}(\{\mu\},\{\lambda\}) \qquad\text{for}\qquad b\neq \ell+1, \\
  H^{+}_{a \ell+1} &=& f^{(a)}(j,\{\lambda\},\{\mu\}),
\end{eqnarray}
with $f^{(a)}(j,\{\lambda\},\{\mu\})$ being the function
(\ref{for2}) defined in the appendix. Exploiting the identities
proven in appendix A we can bring the form factors into the simple
form
\begin{eqnarray}
  F_{\ell}^{+}(j,\{\lambda\}, \{\mu\}) = \frac{\phi_{j-1}(\{\lambda\})}{\phi_{j-1}(\{\mu\})}
 \frac{\prod\limits_{k=1}^{\ell+1}(\mu_k-\xi_j^{-}-s_j \eta)}
  {\prod\limits_{k=1}^{\ell}(\lambda_k-\xi_j^{-}-s_j
  \eta)}\frac{\det \mathcal{C}(\xi_j)}
    {\prod\limits_{1\leq i<j\leq \ell+1} (\mu_j-\mu_i)\prod\limits_{1\leq i<j\leq \ell}
   (\lambda_i-\lambda_j)},
  \label{formminrefin}
\end{eqnarray}
with
\begin{eqnarray}
  \mathcal{C}_{ab} &=& H_{ab}(\{\mu\},\{\lambda\}) \qquad\text{for}\qquad b\neq \ell+1, \\
  \mathcal{C}_{a \ell+1} &=& -i\partial_{\mu_a}p^{(s_j)}(\mu_a
  -\xi_j^{-}). \label{cc}
\end{eqnarray}
\subsection{The form factors of $S_j^{-}$}\indent\\

\noindent In this section we compute the form factors of the
operator $S_j^{-}$ of the XXX spin chain in the spin $s_j$
representation
\begin{eqnarray}
  F^{-}_{\ell}(j,\{\mu\}, \{\lambda\}) &=& \left\langle \psi (\{\mu\})\right|
  S^{-}_{j}\left|\psi (\{\lambda\})\right\rangle.
  \label{formm}
\end{eqnarray}
We can compute these form factors along the same lines of the
previous section. However if we take (\ref{sp}) with (\ref{L1}) as
starting point for our computations we will obtain closed formulae
for (\ref{formm}) which lack the simplicity of
(\ref{formminrefin}). It is in fact more convenient to use
(\ref{Lam2}) instead of (\ref{L1}). Using formula (\ref{Lam2}) and
proceeding as in the previous section it is not difficult to prove
that the form factors (\ref{formm}) are related to
(\ref{formminrefin}) as
\begin{equation}
F_{\ell}^{+}(j,
\{\lambda\},\{\mu\})=\frac{\phi_{j-1}(\{\lambda\})\phi_{j}(\{\lambda\})}
{\phi_{j}(\{\mu\})\phi_{j-1}(\{\mu\})} F_{\ell}^{-}(j,
\{\mu\},\{\lambda\}). \label{dc}
\end{equation}

\section{Conclusions and outlook}
In this paper we have obtained general expressions for the form
factors of all spin operators $\{S^z, S^{\pm}\}$  of the
integrable XXX quantum (higher) spin chain. Our formulae hold for
any spin representation of the operators as well as for any spin
configuration at the remaining sites of the chain. These results
can be  extended to the anisotropic XXZ spin chain as well.

In view of the very general nature of our formulae we expect they
will be useful for the study of a number of interesting physical
systems, whose thermodynamic properties have been already
extensively studied in the literature but whose correlation
functions remain unknown. Amongst these systems alternating spin
chains, first studied in \cite{dVW,deVega:1993sw,alt} and more
recently in \cite{BD}, and impurity systems, such as the Kondo
model \cite{kondo0,kondo1} and the systems considered in
\cite{xxxs,S1,S2}, seem specially interesting examples.

Recent results for the XXX and XXZ spin 1/2 chain show that
determinant formulae for the form factors, similar to those obtained
here, can be successfully employed for numerical computations of
spin-spin dynamical correlation functions of finite chains, which
can match very precisely experimental data \cite{caux1,caux2}. The
form factors computed here might serve for the numerical study of
the correlation functions of impurity systems.

From the analytical point of view, the most natural continuation
of this work is the computation of correlation functions for mixed
spin chains along the lines of \cite{KMT,KMSTa,KMSTb}. Further
interesting generalizations of this would be the study of
dynamical correlation functions in the spirit of \cite{dynamical}
and the computation of correlation functions at finite
temperature, generalizing the program initiated in
\cite{goehmann}.

\paragraph{Acknowledgments:} O.C.A. would like to thank Benjamin Doyon and
Paul P. Martin for their interest in this work, discussions and
the reading of the manuscript, Robert A. Weston for bringing
references
\cite{Idzumi:1993ds,Idzumi:1993sp,Bougourzi:1993mn,Konno:1994gt,Kojima:2005qz}
to her attention and M. Jimbo for one interesting discussion about
the present work. O.C.A. wishes to thank as well the Ecole Normale
Sup\'erieure de Lyon (France) for hospitality. Finally, the
authors acknowledge partial financial support from the EC Network
EUCLID under contract number HPRN-CT-2002-00325.

\appendix
\section{Proof of two identities for the eigenvalues of higher
spin transfer matrices} \label{A} In this appendix we prove two
identities allowing for a compact formula for form factors. Let us
consider two Bethe states $\{\lambda\}=\{\lambda_1, \ldots,
\lambda_\ell\}$ and $\{\mu\}=\{\mu_1, \ldots,
\mu_{\tilde{\ell}}\}$ with $\ell$ and $\tilde{\ell}$ Bethe roots
respectively and define the functions
\begin{eqnarray}
 && g(j, \{\lambda\})=\sum_{k=1}^{2s_j}
 \left[\tau^{({s_j-\frac{k}{2}})}\left(\xi_j + \frac{k\eta}{2}, \{\lambda\}\right)
 \tau^{(\frac{k-1}{2})}\left(\xi_j^{-} + \frac{(k-2s_j)\eta}{2}, \{\lambda\}\right)\right.
  \nonumber \\
 && \left.\times \left[\prod_{p=1}^{{\ell}}
 b^{-1}(\lambda_p-\nu_j(k))-d(\nu_j(k))
 \prod_{p=1}^{{\ell}} b^{-1}(\nu_j(k)-\lambda_p)\right]
 \right],\label{for1}
\end{eqnarray}
and
\begin{eqnarray}
&& f^{(a)}(j, \{\lambda\}, \{\mu\})=
\sum_{k=1}^{2s_j}\left[\tau^{(s_j-\frac{k}{2})}\left(\xi_j+
\frac{k\eta}{2}, \{\lambda\}\right)
  \tau^{(\frac{k-1}{2})}\left(\xi_j^{-} + \frac{(k-2s_j)\eta}{2}, \{\mu\}\right)\right.\label{for2} \\
  &&\times \left.\frac{t(\mu_a,\nu_j(k))}
  {b(\mu_a-\nu_j(k))}\left[ \prod_{p\neq a}b^{-1}(\mu_p-\nu_j(k))-
  d(\nu_j(k)) \prod_{p\neq a}b^{-1}(\nu_j(k)-\mu_p)
  \right]\frac{\prod\limits_{p=1}^{\tilde{\ell}}(\mu_p-\nu_j(k))}
  {\prod\limits_{p=1}^{\ell}(\lambda_p-\nu_j(k))}
 \right].\nonumber
\end{eqnarray}
with $\nu_{j}(k)$ given in (\ref{nu}). The aim of this appendix is
to prove that
\begin{eqnarray}
g(j, \{\lambda\}) &=& 2 s_j
\tau^{(s_j)}(\xi_j,\{\lambda\}), \label{id1}\\
 f^{(a)}(j, \{\lambda\}, \{\mu\}) &=& \frac{\prod\limits_{p=1}^{\tilde{\ell}}
 (\mu_p-\xi_{j}^{-}+s_j\eta)}
 {\prod\limits_{p=1}^{{\ell}}(\lambda_p-\xi_{j}^{-}- s_j\eta)}\sum_{k=1}^{2s_j} t(\mu_a,\nu_{j}(k)),\label{id2}
\end{eqnarray}
with,
\begin{equation}
\sum_{k=1}^{2 s_j}
  t(\mu_a, \nu_{j}(k))=-i\partial_{\mu_a} p^{(s_j)}(\mu_a-\xi_j^{-}),
\end{equation}
and
\begin{equation}\label{pp}
p^{(s_j)}(\lambda)=i \log\left[\frac{\lambda-s_j \eta}{\lambda +
s_j \eta}\right].
\end{equation}
is the momentum of a spin $s_j$ pseudo-particle of ``rapidity"
$\lambda$. In particular, the identities above imply
\begin{eqnarray}
 f^{(a)}(j, \{\lambda\}, \{\lambda\}) =\tau^{(s_j)}(\xi_j,\{\lambda\})
\left[ -i\partial_{\lambda_a}
p^{(s_j)}(\lambda_a-\xi_j^{-})\right],
\end{eqnarray}
\indent \newline Let us commence our proof by writing down the
explicit expression of the eigenvalues following from
eq.(\ref{ex})
\begin{eqnarray}
   && \tau^{({s_j-\frac{k}{2}})}\left(\xi_j + \frac{k\eta}{2}, \{\lambda\}\right)
   = \sum_{\alpha=0}^{2s_j-k}\left[
    C_{\alpha}^{(s_j-\frac{k}{2})}\left(\xi_j + \frac{k\eta}{2}\right) \right.\nonumber \\
    && \left.\times \prod_{p=1}^{\ell}
    \frac{(\xi_j^{+} + s_j\eta-\lambda_p)
   (\nu_{j}(k)-\lambda_p )}{(\nu_{j}(k) + (\alpha+1 )\eta-\lambda_p)
    (\nu_{j}(k) + \alpha\eta-\lambda_p)}\right], \label{a2}\\
 && \tau^{(\frac{k-1}{2})}\left(\xi_j^{-} + \frac{(k-2s_j)\eta}{2},
 \{\mu\}\right)=
  \sum_{\beta=0}^{k-1}\left[
    C_{\beta}^{(\frac{k-1}{2})}\left(\xi_j^{-} + \frac{(k-2s_j)\eta}{2}\right) \right. \nonumber \\
    &&\left.\times \prod_{p=1}^{\tilde{\ell}} \frac{(\nu_{j}(k)-\mu_p)
    (\xi_j^{-}- s_j\eta-\mu_p)}{(\xi_j^{+}- (s_j-\beta)\eta-\mu_p)
    (\xi_j^{-}- (s_j-\beta)\eta-\mu_p)}\right],\label{b}
\end{eqnarray}
whose product enters the two formulae we want to prove. Before we
insert these formulae in  (\ref{for1})-(\ref{for2}) it is useful
to note that in (\ref{a2}) the only non vanishing term is the one
corresponding to $\alpha=2s_j-k$. It is easy to see that any other
terms will contain a factor $d(\xi_j^{-}+s_j\eta)=0$ and never
contain the singular factor $d(\xi_j^{-}-s_j\eta)$. Likewise it is
straightforward to argue that such factors will never appear in
$C_{\beta}^{((k-1)/2)}(\xi_j^{-} + {(k-2s_j)\eta}/{2})$ for any of
the allowed values of $\beta$ and $k$, and therefore each term in
the $\beta$-sum is non-vanishing and non-singular. Therefore
\begin{eqnarray}
 && \tau^{({s_j-\frac{k}{2}})}\left(\xi_j + {k\eta}/{2}, \{\lambda\}\right)
 \tau^{(\frac{k-1}{2})}\left(\xi_j^{-} + {(k-2s_j)\eta}/{2}, \{\mu\}\right)
  \nonumber \\
  && = \frac{\prod\limits_{p=1}^{\tilde{\ell}}(\mu_p -\xi_j^{-} +s_j \eta)}
  {\prod\limits_{p=1}^{{\ell}}(\lambda_p -\xi_j^{-} - s_j \eta)}
  \sum_{\beta=0}^{k-1} C_{\beta}^{(\frac{k-1}{2})}(\xi_j^{-} + {(k-2s_j)\eta}/{2})
  \nonumber \\
  && \times
   \frac{\left[\prod\limits_{p=1}^{\tilde{\ell}} (\mu_p-\nu_j(k))\right]\left[
   \prod\limits_{p=1}^{\ell}(\lambda_p-\nu_j(k))\right]}
  {\prod\limits_{p=1}^{\tilde{\ell}} (\mu_p-\xi_j^{+}+(s_j-\beta)\eta)
    (\mu_p-\xi_j^{-}+ (s_j-\beta)\eta)},
\end{eqnarray}
where we used the second property stated in (\ref{C}), i.e.
$C_{2s_j-k}^{(s_j-k/2)}(\xi_j + k\eta/2)=1$. Using these formulae,
the equalities (\ref{for1}) and (\ref{for2}) are equivalent to
proving :
\begin{eqnarray}
 &&\sum_{k=1}^{2s_j}\sum_{\beta=0}^{k-1} \left[C_{\beta}^{(\frac{k-1}{2})}\left(\xi_j^{-} + \frac{{(k-2s_j)\eta}}{2}\right)
 \prod_{p=1}^{\ell} \frac{(\lambda_p-\nu_j(k))^{2}}
  {h(\lambda_p-\xi_j^{+}+(s_j-\beta)\eta)}\right.\nonumber\\
  &&\times \left.\left[\prod_{p=1}^{\ell}
b^{-1}(\lambda_p-\nu_j(k))-d(\nu_j(k))
  \prod_{p=1}^{\ell}
  b^{-1}(\nu_j(k)-\lambda_p)\right]\right]=2s_j,\label{fa}
\end{eqnarray}
and
\begin{eqnarray}
&&\sum_{k=1}^{2s_j}\sum_{\beta=0}^{k-1}
\left[C_{\beta}^{(\frac{k-1}{2})}\left(\xi_j^{-} +
\frac{{(k-2s_j)\eta}}{2}\right)
 \prod_{p=1}^{\tilde{\ell}} \frac{(\mu_p-\nu_j(k))^{2}}
  {h(\mu_p-\xi_j^{+}+(s_j-\beta)\eta)}\right. \nonumber \\
  &&\times \left[ \prod_{p\neq a}b^{-1}(\mu_p-\nu_j(k))-
  d(\nu_j(k)) \prod_{p\neq a}b^{-1}(\nu_j(k)-\mu_p)
  \right]  \nonumber \\
  &&\left.\times \frac{t(\mu_a,\nu_j(k))}{b(\mu_a-\nu_j(k))}
 \right]= -i\partial_{\lambda_a} p^{(s_j)}(\lambda_a-\xi_j^{-}),\label{fb}
\end{eqnarray}
respectively, with
\begin{equation}\label{h}
    h(\lambda):={(\lambda)}{(\lambda + \eta)}.
\end{equation}
We commence by establishing (\ref{fa}). Let us consider those
terms in (\ref{fa}) which do not contain any factors proportional
to the function $d(x)$. The only terms of this kind correspond to
taking $\beta=k-1$ and only the first contribution on the second
line of (\ref{fa}), that is
\begin{eqnarray}
 &&\!\!\!\!\!\!\!\!\!\!\! \sum_{k=1}^{2s_j}
 \prod_{p=1}^{\ell} \frac{(\lambda_p-\xi_j^{-}+ (s_j-k)\eta)(\lambda_p-\xi_j^{-}-(k-s_j-1)\eta)}
  {(\lambda_p-\xi_j^{+}+(s_j-k+1)\eta)
  (\lambda_p-\xi_j^{-}+ (s_j-k+1)\eta)}=2s_j,
\end{eqnarray}
therefore it remains to prove that all remaining terms cancel each
other. Those remaining terms are
\begin{eqnarray}
 &&\sum_{k=2}^{2s_j}\sum_{\beta=0}^{k-2}
 C_{\beta}^{(\frac{k-1}{2})}\left(\xi_j^{-} + \frac{(k-2s_j)\eta}{2}\right)
 \prod_{p=1}^{\ell} \frac{h(\lambda_p-\nu_j(k))}
  {h(\lambda_p-\xi_j^{+}+(s_j-\beta)\eta)}\nonumber \\
  &&-
  \sum_{k=1}^{2s_j}\sum_{\beta=0}^{k-1} C_{\beta}^{(\frac{k}{2})}\left(\xi_j + \frac{(k-2s_j)\eta}{2}\right)
 \prod_{p=1}^{\ell} \frac{h(\lambda_p-\nu_j(k)-\eta)}
  {h(\lambda_p-\xi_j^{+}+(s_j-\beta)\eta)}, \label{ss}
\end{eqnarray}
where we have used the identity
\begin{equation}
   C_{\beta}^{(\frac{k-1}{2})}\left(\xi_j^{-} + \frac{{(k-2s_j)\eta}}{2}\right)\, d(\nu_j(k))
   =  C_{\beta}^{(\frac{k}{2})}\left(\xi_j + \frac{{(k-2s_j)\eta}}{2}\right).
\end{equation}
Replacing the sum over $k$ by a sum over $k^{\prime}=k-1$ in the
first line of (\ref{ss}) we see that both terms are essentially
identical but for the extra term $k=2s_j$ contributing in the
second sum. However such term is vanishing since
$C_{\beta}^{(s_j)}(\xi_j^{-} + {(k-2s_j)\eta}/{2})=0$ and
therefore we have proven that indeed all terms containing the
function $d(x)$ in (\ref{fa}) cancel each other. With this we have
established (\ref{fa}) and therefore (\ref{for1}). The proof of
the identity (\ref{fb}) can be carried out by following exactly
the same steps, namely considering separately those terms which
contain $d$-functions and those which do not.

\section{An alternative formula for the solution of the inverse scattering problem for XXX spin chains}
\label{inverse} The solution of the inverse scattering problem for
XXX spin chains and arbitrary spin representations was given in
\cite{JM1}. This solution was recalled in formulae (\ref{sp}) and
(\ref{L1}) and employed in order to obtain closed expressions for
higher spin form factors of the operators $S^{z}$ and $S^{+}$.
However we have observed that in some cases the form factor formulae
obtained from (\ref{L1}) simplify considerably (thanks to the
identities proven in appendix A), whereas in other cases such
simplifications do not occur. For example, a simple formula for all
form factors of the operator $S^{+}$ can be obtained form
(\ref{sp})-(\ref{L1}), but the formula we would obtain for the form
factors of $S^{-}$ from (\ref{L1}) is much more complicated. This
asymmetry seems rather unnatural and one may suppose that the
formulae for the $S^{-}$ form factors can be further simplified and
recasted into analogous formulae as those for the $S^{+}$ form
factors. This should be doable by using properties of determinants
as well as the Bethe ansatz equations; however it turns out to be a
rather non-trivial proof. An alternative way of finding simpler
formulae for the form factors of $S^{-}$ is to start with a
different (but equivalent) reconstruction formula, that is
(\ref{Lam2}). In this appendix we will show that (\ref{L1}) and
(\ref{Lam2}) are indeed equivalent. The key tool are once again the
fusion relations (\ref{dttd}) and one particular property of these
noticed in \cite{KRS}.
\subsection{An alternative version of the fusion relations}\ \\

\noindent Let us start by decomposing (\ref{dttd}) into the
following two equations
\begin{eqnarray}
P^{+}_{a_1 \{a\}}T_{a_1}^{(\frac{1}{2})}(x^{-}+s\eta)
T_{\{a\}}^{(s-\frac{1}{2})}(x^{-})P^{+}_{a_1 \{a\}} \!\!\!&=&\!
\!\!T_{\langle a_1
\{a\} \rangle}^{(s)}(x), \label{c1} \\
P^{-}_{a_1 \{a\}}T_{a_1}^{(\frac{1}{2})}(x^{-}+s\eta)
T_{\{a\}}^{(s-\frac{1}{2})}(x^{-})P^{-}_{a_1 \{a\}} \!\!\!&=&\!
\!\! \chi(x+(s-1)\eta) T_{( a_1 \{a\} )}^{(s-1)}(x-\eta)
\label{c2}.
\end{eqnarray}
With respect to (\ref{dttd}) we have slightly changed our notation
and introduced the index $\{a\}$ to indicate that the space
$V_{\{a\}}^{(s-1/2)} \sim \mathbb{C}^{2s-1}$ is isomorphic to the
space $ V_{\left\langle a_{2} \ldots
a_{2s}\right\rangle}^{(s-1/2)}$ resulting from the fusion of
$2s-1$ spin 1/2 quantum spaces
\begin{equation}
V_{a_j}^{(\frac{1}{2})} \sim \mathbb{C}^{2},\qquad j=2,\ldots, 2s.
\end{equation}
We can now employ fusion successively in order to express
(\ref{c1}) solely in terms of spin 1/2 quantum monodromy matrices.
By doing so, we obtain the following expression
\begin{eqnarray}
P^{+}_{a_1 \ldots a_{2s}}
\prod_{j=1}^{2s}T_{a_j}^{(\frac{1}{2})}(x^{+}+(s-j)\eta)
P^{+}_{a_1 \ldots a_{2s}} =T_{\langle a_1 \ldots a_{2s}
\rangle}^{(s)}(x). \label{c11}
\end{eqnarray}
It was proven in  \cite{KRS} that,  for XXX spin chains, the
operator $P^{+}_{a_1 \ldots a_{2s}}$ is a complete symmetrizer on
all indices $a_1, \ldots, a_{2s}$. Let us now consider the formula
(\ref{c2}). We can also express this relation entirely in terms of
spin 1/2 monodromy matrices
\begin{eqnarray}
P^{-}_{a_1 \{a\}} P^{+}_{a_2 \ldots a_{2s}}
\prod_{j=1}^{2s}T_{a_j}^{(\frac{1}{2})}(x^{+}+(s-j+1)\eta)
P^{+}_{a_2 \ldots a_{2s}}P^{-}_{a_1 \{a\}} = \chi(x+ s\eta) T_{(
a_1 \{a\}) }^{(s-1)}(x) \label{c22}.
\end{eqnarray}
As before the operator $P^{+}_{a_2 \ldots a_{2s}}$ is completely
symmetric on the indices $\{a_2 \ldots a_{2s}\}$. On the other
hand the operator $P^{-}_{a_1 \{a\} }$ is antisymmetric on the
index $a_1$ (see \cite{KRS}). The identities
(\ref{c11})-(\ref{c22}) can be further manipulated by employing
the following property of the projectors $P^{\pm}$ which was
stated in \cite{KRS}
\begin{equation}\label{proj}
    P^{\pm}_{a_1 \ldots a_n} R^{(\frac{1}{2},\frac{1}{2})}_{a_1 a_{n+1}}(\pm n
    \eta) P^{\pm}_{a_2 \ldots a_{n+1}}= \pm P^{\pm}_{a_1 \ldots
    a_{n+1}}.
\end{equation}
This property  allowed the authors of \cite{KRS} to prove that
\begin{equation}
P^{+}_{a_1 \ldots a_{2s}}
\prod_{j=1}^{2s}T_{a_j}^{(\frac{{1}}{2})}(x^{+}+(s-j)\eta)
P^{+}_{a_1 \ldots a_{2s}}=P^{+}_{a_1 \ldots a_{2s}}
\prod_{j=1}^{2s}T_{a_j}^{(\frac{{1}}{2})}(x^{-}-(s-j)\eta)
P^{+}_{a_1 \ldots a_{2s}}, \label{i1}
\end{equation}
and
\begin{eqnarray}
&&P^{-}_{a_1 \{a\}} P^{+}_{a_2 \ldots a_{2s}}
\prod_{j=1}^{2s}T_{a_j}^{({1}/{2})}(x^{+}+(s-j)\eta) P^{+}_{a_2
\ldots a_{2s}} P^{-}_{a_1 \{a\}} \nonumber \\
&&= P^{-}_{a_1 \{a\}} P^{+}_{a_2 \ldots a_{2s}}
\prod_{j=1}^{2s}T_{a_j}^{({1}/{2})}(x^{-}-(s-j)\eta) P^{+}_{a_2
\ldots a_{2s}} P^{-}_{a_1 \{a\}}. \label{i2}
\end{eqnarray}
The equalities (\ref{i1})-(\ref{i2}) imply that the fusion relation
(\ref{dttd}) still holds if in the arguments on the r.h.s. $\eta$ is
replaced by $-\eta$, that is
\begin{eqnarray}
P_{12} T_{1}^{(\frac{1}{2})}(x^{+}-s \eta)
T_{2}^{(s-\frac{{1}}{2})}(x^{+}) P_{12} =\left(\begin{array}{cc}
T_{\langle12 \rangle }^{(s)}(x) & 0 \\
* & \chi(x+(s-1)\eta) T_{( 12 )}^{(s-1)}(x-\eta) \\
\end{array}
\right). \label{dttd2}
\end{eqnarray}
This equivalence was stated in equation (18) of \cite{KRS} for
$R$-matrices, but it is easy to prove that it implies a similar
property of the monodromy matrices.

\paragraph{The $s=1$ case:} As an example, let us prove (\ref{i1}) for $s=1$.
In this case (\ref{c11}) becomes simply
\begin{eqnarray}
P^{+}_{a_1 a_2}
T_{a_1}^{(\frac{{1}}{2})}(x^{+})T_{a_2}^{(\frac{1}{2})}(x^{-})
P^{+}_{a_1 a_{2}} &=& T_{\langle a_1 a_{2} \rangle}^{(1)}(x),
\end{eqnarray}
and the relation (\ref{proj}) can be written as
\begin{equation}
   R^{(\frac{1}{2},\frac{1}{2})}_{a_1 a_{2}}(\pm n
    \eta) = \pm P^{\pm}_{a_1
    a_{2}},
\end{equation}
for $n=1$. The $RTT$-relations
\begin{equation}
R_{a_1
a_2}^{(\frac{1}{2},\frac{1}{2})}(\lambda)T_{a_1}^{(\frac{1}{2})}(\lambda
+ \mu) T_{a_2}^{(\frac{1}{2})}(\mu)=
T_{a_2}^{(\frac{1}{2})}(\mu)T_{a_1}^{(\frac{1}{2})}(\lambda +
\mu)R_{a_1 a_2}^{(\frac{1}{2},\frac{1}{2})}(\lambda), \label{rtt}
\end{equation}
at $\lambda=\eta$ and $\mu=x^{-}$ imply therefore
\begin{equation}
P_{a_1 a_2}^{+} T_{a_1}^{(\frac{1}{2})}(x^{+})
T_{a_2}^{(\frac{1}{2})}(x^{-})=
T_{a_2}^{(\frac{1}{2})}(x^{-})T_{a_1}^{(\frac{1}{2})}(x^{+})P_{a_1
a_2}^{+}. \label{rtt2}
\end{equation}
Multiplying (\ref{rtt2}) from the right and from the left by
$P_{a_1 a_2}^{+}$ we obtain
\begin{eqnarray}
P^{+}_{a_1 a_2}
T_{a_1}^{(\frac{1}{2})}(x^{+})T_{a_2}^{(\frac{1}{2})}(x^{-})
P^{+}_{a_1 a_{2}} = P^{+}_{a_1 a_2}
T_{a_2}^{(\frac{1}{2})}(x^{-})T_{a_1}^{(\frac{1}{2})}(x^{+})
P^{+}_{a_1 a_{2}}, \label{rtt3}
\end{eqnarray}
which is precisely (\ref{i1}) for $s=1$.
\subsection{An alternative formula for the solution of the inverse problem}\ \\

\noindent In order to solve the inverse problem we need to find a
systematic way of computing the traces
\begin{equation}
 \Lambda_{\alpha}^{(s_j)}:=\text{Tr}_{0}\left(S_{0}^{\alpha} T^{(s_j)}_{0; 1\ldots
 N}(\xi_j)\right),\qquad \alpha=\pm ,z. \label{LL}
\end{equation}
To do that, the key idea \cite{JM1} is to use the fusion relations
(\ref{dttd2}) in order to obtain recursive formulae similar to
(\ref{ide1}) for (\ref{LL}). Recall that for the general XXX chain
in finite dimensional representations, the co-product for local
spin operator is trivial and therefore we can write the spin
matrices in the two following ways
\begin{equation}
    S_{12}^{\alpha}=I_{1} \otimes
    S_{2}^{\alpha} +S_{1}^{\alpha} \otimes
    I_{2} \simeq S_{\langle 12\rangle}^{\alpha} \oplus
    S_{(12)}^{\alpha}, \qquad \alpha=\pm,z.
\end{equation}
Multiplying (\ref{dttd2}) by $S_{12}^{\alpha}$ and taking
thereafter the trace over the spaces $1$ and $2$ on the l.h.s. and
over the fused spaces $\langle 12 \rangle$ and
 $( 12)$ on the r.h.s. we obtain the following relations
 \begin{eqnarray}
 && t^{(\frac{1}{2})}(x^{+}-s\eta)
\Lambda_{\alpha}^{(s-\frac{1}{2})}(x^{+}) +
\Lambda_{\alpha}^{(\frac{1}{2})}(x^{+}-s\eta
)t^{(s-\frac{1}{2})}(x^{+}) \nonumber \\
&& =\Lambda_{\alpha}^{(s)}(x) + \chi(x+(s-1)\eta)
\Lambda_{\alpha}^{(s-1)}(x-\eta). \label{rec}
 \end{eqnarray}
 In addition, taking directly the trace on (\ref{dttd2}) we obtain
 also new fusion relations for the traces of the transfer matrices
\begin{eqnarray}
   && t^{({1}/{2})}(x^{+}-s\eta)
t^{(s-1/2)}(x^{+})  =t^{(s)}(x) + \chi(x+(s-1)\eta)
t^{(s-1)}(x-\eta). \label{fr}
 \end{eqnarray}
The recursive relations (\ref{rec}) are solved by
\begin{eqnarray}
  \Lambda_{\alpha}^{(s)}(u)=
  \sum_{k=1}^{2 s} t^{(s-\frac{k}{2})}\left(u - \frac{k\eta}{2}\right)
  \Lambda_{\alpha}^{(\frac{1}{2})}(u^{+}-(k-s)\eta)
 t^{(\frac{k-1}{2})}\left(u^{+} -
  \frac{(k-2s)\eta}{2}\right).\label{L2}
\end{eqnarray}
This can be proven by substituting $\Lambda_{\alpha}^{(s-1)}$ and
$\Lambda_{\alpha}^{(s-1/2)}$ in (\ref{rec}) by the corresponding
formulae from (\ref{L2}) and using thereafter the fusion relations
(\ref{fr}). This is exactly the same solution (\ref{L1}) up to the
replacement $\eta \rightarrow -\eta$ in the arguments. By
introducing a new index $p=2s-k +1$ we can rewrite (\ref{L2}) as
\begin{eqnarray}
  \Lambda_{\alpha}^{(s)}(u)=
  \sum_{p=1}^{2 s} t^{(\frac{p-1}{2})}\left(u^{-} + \frac{(p-2s)\eta}{2}\right)
  \Lambda_{\alpha}^{(\frac{1}{2})}(u^{-}+(p-s)\eta)
 t^{(s-\frac{p}{2})}\left(u +
  \frac{p\eta}{2}\right),\label{L22}
\end{eqnarray}
which is identical to (\ref{L1}) up to the exchange of the
transfer matrices.

\small
%\bibliography{Ref}

\end{document}